\documentclass[twocolumn,showpacs,aps,prl,superscriptaddress]{revtex4}

\usepackage{xspace}
\usepackage{graphicx}
\usepackage{dcolumn}
\usepackage{amsmath}
\usepackage{epsfig}

\input babarsym.tex

\newcommand{\eett}   {\ensuremath{e^+e^- \to \tautau}\xspace}

\newcommand{\roots}        {\ensuremath{\sqrt{s}}\xspace}

\newcommand{\Vusvalue} {\ensuremath{0.2255 \pm 0.0024}\xspace}
\newcommand{\Vusunit} {\ensuremath{0.2255 \pm 0.0010}\xspace}

\newcommand{\gmgevalue} {\ensuremath{1.0036 \pm 0.0020}\xspace}

\newcommand{\gtgmpvalue} {\ensuremath{0.9856 \pm 0.0057}\xspace}
\newcommand{\gtgmkvalue} {\ensuremath{0.9827 \pm 0.0086}\xspace}
\newcommand{\gtgmhvalue} {\ensuremath{0.9850 \pm 0.0054}\xspace}

\def\SM             {Standard Model\xspace}

\newcommand{\tauknu}   {\ensuremath{ \tau^{-} \to K^{-} \nut}\xspace}
\newcommand{\taupinu}   {\ensuremath{ \tau^{-} \to \pi^{-} \nut}\xspace}
\newcommand{\tauhnunc}   {\ensuremath{ \tau \to h \nut}\xspace}

\newcommand{\tauenu}   {\ensuremath{ \tau^{-} \to e^{-} \nueb \nut}\xspace}
\newcommand{\taumunu}  {\ensuremath{ \tau^{-} \to \mu^{-} \numb \nut}\xspace}

\newcommand{\BFtautoknu}    {\ensuremath{\BR(\tauknu)}\xspace}					
\newcommand{\BFtautopinu}    {\ensuremath{\BR(\taupinu)}\xspace}
\newcommand{\BFtautohnunc}    {\ensuremath{\BR(\tauhnunc)}\xspace}

\newcommand{\BFhmutwonc}    {\ensuremath{\BR(h \to \mu \num)}\xspace}

\newcommand{\BRtautoknu}    {\ensuremath{ \frac{\BR(\tauknu)}{\BR(\tauenu)} }\xspace}			
\newcommand{\BRtautopinu}    {\ensuremath{\frac{\BR(\taupinu)}{\BR(\tauenu)} }\xspace}			
\newcommand{\BRtautomunu}    {\ensuremath{\frac{\BR(\taumunu)}{\BR(\tauenu)} }\xspace}

\newcommand{\BFknu}   {\ensuremath{  (0.692 \pm 0.006 \pm 0.010 )\% }\xspace}
\newcommand{\BFpinu}   {\ensuremath{ (10.59 \pm 0.03 \pm 0.11)\% }\xspace}

\newcommand{\BFmunu}   {\ensuremath{ (17.46 \pm 0.03 \pm 0.08)\%  }\xspace}

\newcommand{\BRknu}   {\ensuremath{ (0.03882 \pm 0.00032 \pm 0.00057)}\xspace}
\newcommand{\BRpinu}   {\ensuremath{(0.5945 \pm 0.0014  \pm 0.0061)}\xspace}
\newcommand{\BRhnu}   {\ensuremath{ (0.6333 \pm 0.0014 \pm 0.0061)}\xspace}

\newcommand{\BRmunu}   {\ensuremath{(0.9796 \pm 0.0016  \pm 0.0036)}\xspace}
\newcommand{\BRkpi}   {\ensuremath{(0.06531 \pm 0.00056  \pm 0.00093)}\xspace}

\newcommand{\Effknu}   {\ensuremath{ 0.330\% }\xspace}
\newcommand{\Effpinu}  {\ensuremath{ 0.324\% }\xspace}
\newcommand{\Effenu}   {\ensuremath{ (0.589\pm0.010)\% }\xspace}
\newcommand{\Effmunu}   {\ensuremath{0.485\%  }\xspace}

\newcommand{\Purknu}   {\ensuremath{ 76.6\% }\xspace}
\newcommand{\Purpinu}  {\ensuremath{ 78.7\% }\xspace}
\newcommand{\Purenu}   {\ensuremath{ (99.69\pm0.06)\% }\xspace}
\newcommand{\Purmunu}   {\ensuremath{ 97.3\%} \xspace}

\newcommand{\NDataknu}   {\ensuremath{ 25123 }\xspace}
\newcommand{\NDatapinu}   {\ensuremath{369091 }\xspace}
\newcommand{\NDataenu}   {\ensuremath{ 884426 }\xspace}
\newcommand{\NDatamunu}   {\ensuremath{731102 }\xspace}

\newcommand{\ntaupair}         {\ensuremath{4.29\times 10^8}\xspace}

\newcommand{\lumi}             {\ensuremath{467\invfb}\xspace}

\def\kk       {\mbox{\tt KK}\xspace}
\def\tauola     {\mbox{\tt TAUOLA}\xspace}
\def\photos     {\mbox{\tt PHOTOS}\xspace}

\newcommand{\gevccgevcc}{\ensuremath{{\mathrm{\,Ge\kern -0.1em V^2\!/}c^4}}\xspace}

\newcommand{\evcc}{\ensuremath{{\mathrm{\,e\kern -0.1em V\!/}c^2}}\xspace}

\newcommand{\CM} {\mbox{CM}\xspace}

\newcommand{\THETAMISSCM} {\ensuremath{\theta_{\rm{miss}}^{\rm{CM}}}}

\newcommand{\BABARPubYear}     {09}
\newcommand{\BABARPubNumber}  {018}
\newcommand{\SLACPubNumber} {13846}
\newcommand{\LANLNumber}  {0912.0242 [hep-ex]}

\def\figurebox#1#2#3{%
    \def\arg{#3}%
    \ifx\arg\empty
    {\hfill\vbox{\hsize#2\hrule\hbox to #2{\vrule\hfill\vbox to #1{\hsize#2\vfill}\vrule}\hrule}\hfill}%
    \else
    {\hfill\epsfbox{#3}\hfill}%
    \fi}

\begin{document}

\preprint{\babar-PUB-\BABARPubYear/\BABARPubNumber} 
\preprint{SLAC-PUB-\SLACPubNumber} 
\preprint{\LANLNumber}    

\begin{flushleft}
\babar-PUB-\BABARPubYear/\BABARPubNumber\\
SLAC-PUB-\SLACPubNumber\\
arXiv:\LANLNumber \\[10mm]    
\end{flushleft}

\vspace*{-.5cm}

\title{
{\large \bf \boldmath
Measurements of Charged Current Lepton Universality and $|V_{us}|$
using \\ Tau Lepton Decays to e$^-~\nueb\nut$, $\mu^-~\numb\nut$, $\pi^-\nut$, and $K^-\nut$}}

%
\author{B.~Aubert}
\author{Y.~Karyotakis}
\author{J.~P.~Lees}
\author{V.~Poireau}
\author{E.~Prencipe}
\author{X.~Prudent}
\author{V.~Tisserand}
\affiliation{Laboratoire d'Annecy-le-Vieux de Physique des Particules (LAPP), Universit\'e de Savoie, CNRS/IN2P3,  F-74941 Annecy-Le-Vieux, France}
\author{J.~Garra~Tico}
\author{E.~Grauges}
\affiliation{Universitat de Barcelona, Facultat de Fisica, Departament ECM, E-08028 Barcelona, Spain }
\author{M.~Martinelli$^{ab}$}
\author{A.~Palano$^{ab}$ }
\author{M.~Pappagallo$^{ab}$ }
\affiliation{INFN Sezione di Bari$^{a}$; Dipartimento di Fisica, Universit\`a di Bari$^{b}$, I-70126 Bari, Italy }
% ...
\collaboration{The \babar\ Collaboration}
\noaffiliation

\date{\today}

\begin{abstract}
Using \lumi of \epem annihilation data collected with the \babar\ detector,
we measure \BRtautomunu=\BRmunu, \BRtautopinu=\BRpinu, and \BRtautoknu=\BRknu, 
where the uncertainties are statistical and systematic, respectively.
From these precision $\tau$ measurements,
we test the \SM assumption of $\mu$-$e$ and $\tau$-$\mu$ charge current lepton universality and
provide  determinations of $\Vus$ experimentally 
independent of the decay of a kaon.
\end{abstract}


\maketitle

Decays of the $\tau$ lepton to a single charged particle and neutrino(s)
probe the \SM (SM) predictions of charged current lepton universality
 and the unitarity relation of the first row of the
 Cabibbo-Kobayashi-Maskawa (CKM) quark mixing matrix~\cite{CKM}.
Previous measurements of universality~\cite{Pich:2008ni,PDG},  
 expressible in terms of the coupling strength ($g_\ell$) of 
lepton of flavor $\ell$ to the charged gauge boson of the  electroweak interaction
are in agreement with the SM where $g_{\tau}/g_{\mu}=g_{\mu}/g_{e}=1$.
Similarly, kaon decay measurements~\cite{PDG,:2008ct} sensitive to  \Vus,
the relative weak coupling between up and strange quarks,
yield a value consistent with unitarity ($\Vud^2+\Vus^2+\Vub^2=1$) 
where nuclear beta decays provide \Vud~\cite{Vud} and \Vub is negligible~\cite{PDG}.
However, new physics that couples primarily to the third generation could be revealed through
deviations from the SM in precision universality and \Vus  measurements involving the $\tau$. 
Significant deviations of this nature are unambiguous signatures of new physics that provide
crucial but complimentary information to the direct searches for Higgs bosons~\cite{Logan:2009uf}
and other new physics models with e.g. lepto-quarks~\cite{Dorsner:2009cu},
heavy gauge $W^\prime$ or $Z^\prime$ bosons, heavy quarks or leptons,
 compositeness or extra dimensions~\cite{Loinaz:2002ep}.

Recent measurements of the sum of strange $\tau$ branching fractions
interpreted in the framework of the Operator Product Expansion (OPE) and finite energy sum rules 
yield a value of \Vus that is approximately three standard deviations ($\sigma$) lower than expectations
 from CKM unitarity~\cite{Maltman:2008ib}. 
 This paper addresses both experimental and theoretical aspects of this question by 
providing the first precision measurements of 
$R_{K}\equiv\BRtautoknu$~\cite{conjugate} and
$R_{K/\pi}\equiv\frac{\BFtautoknu}{\BFtautopinu}$  enabled by
the unique combination of a very large $\tau$ sample with particle momenta amenable to 
particle identification using Cherenkov radiation. 
By using values of the meson decay constants from lattice QCD~\cite{Follana:2007uv},
we provide two precision determinations of \Vus from $\tau$ decays independent of the OPE framework.
We also report on new measurements of 
$R_{\pi}\equiv$\BRtautopinu and $R_{\mu}\equiv$\BRtautomunu.
 $R_{\mu}$ provides an improved measurement of  
$g_{\mu}/g_{e}$ whereas $R_{\pi}$ and $R_{K}$, when compared to the muonic branching fractions of the pion
and kaon, yield improved measurements of $g_{\tau}/g_{\mu}$ involving pseudoscalar mesons.

The data sample corresponds to an integrated luminosity of $\mathcal{L}$ = \lumi
recorded at an $e^+e^-$ center-of-mass (\CM) energy (\roots) near 10.58\gev
and was collected with the \babar\ detector at the SLAC \pep2  \epem storage rings.
With a luminosity-weighted average cross-section of $\sigma_{\eett}=(0.919\pm0.003)$ nb~\cite{Banerjee:2007is,kk}, 
this corresponds to the production of \ntaupair $\tau$-pair events.
The \babar\ detector~\cite{detector} is composed of a silicon vertex tracker, drift chamber (DCH),
ring-imaging Cherenkov detector (DIRC), and 
 electromagnetic calorimeter (EMC), all contained in a 1.5-T solenoid.
The iron flux return for the solenoid is instrumented (IFR) to identify muons.

Tau-pair events are simulated with the \kk Monte Carlo (MC) generator~\cite{kk}, which
includes higher-order radiative corrections.
We simulate $\tau$ decays with \tauola~\cite{tauola} and \photos~\cite{photos}
using measured branching fractions~\cite{PDG}.
The detector response is simulated with \mbox{\tt GEANT4}~\cite{geant}. 
Simulated events for signal as well as background processes~\cite{kk,tauola,photos,Lange:2001uf,Sjostrand:1995iq}
are reconstructed in the same manner as data.
The MC samples are used for selection optimization, control sample studies, and systematic error studies.
The number of simulated non-signal events is comparable to the number expected in the data,
with the exception of Bhabha and two-photon events, which are not simulated but which data studies
show to be negligible.


We study $e^+e^-\ra\tau^+\tau^-$ events with the $\tau^-$ decaying via
$\tauenu$, $\taumunu$, $\taupinu$ or $\tauknu$ modes
and the $\tau^{+}$ decaying via a $\tau^{+} \to \pi^{+}\pi^{+}\pi^{-}\nutb$ tagging channel
with the selection criteria optimized to minimize the combined statistical and 
systematic uncertainties~\cite{Nugent}.
The number of signal events 
for decay modes $i$ =  $\{e, \mu, \pi, K\}$ = $\{e^-\nueb\nut, \mu^-\numb \nut, \pi^-\nut, K^-\nut\}$ are
$N_{i}^{\rm{S}} =  \mathbf{\cal E}^{-1}_{i}\left( N_{i}^{\rm{D}}-N_{i}^{\rm{B}}\right)$
where $\mathbf{\cal E}_{i}$ is the efficiency (including $\BR(\tau^-\to\pi^{-}\pi^{-}\pi^{+}\nu_{\tau})=(8.85\pm0.13)\%$~\cite{PDG}),
$N^{\rm{D}}_i$ the number of selected data events, and
$N^{\rm{B}}_i$ the estimated number of background events for the $i^{th}$ mode.

We measure the ratios $R_i = N_{i}^{\rm{S}}/N_{e}^{\rm{S}}$ 
which  normalizes to the most precisely known relevant SM process available,
and in which several common sources of systematic uncertainity cancel.
$N^{\rm{D}}_i$ are multiplied with reproducible random numbers 
until all efficency and uncertainity estimates are finalized.
Once unblinded, we use the values of the three branching ratios to
update world averages of the branching fractions, which we then use to 
 recalculate the backgrounds for our final results.

Events with a net charge of zero and with four well-reconstructed tracks 
not originating from the conversion of a photon in the detector material are selected. 
For good particle identification, each track is required to be within the acceptance of the DIRC and EMC, 
and have a transverse momentum greater than 0.25\gev to ensure that it reaches the DIRC.
The plane normal to the thrust axis 
 divides the event into hemispheres in the \CM frame. 
The ``signal'' hemisphere contains a single track and the ``tag'' hemisphere the other three tracks.

Each tag hemisphere track is
 required to be consistent with being a pion and 
the energy  deposited in the EMC  unassociated with any tracks in this hemisphere
 is required to be less than 0.20\gev.
Also, events that contain track pairs consistent with coming from a $K_{S}^{0}$ are vetoed.

The signal track momentum is required to lie between 1 and 4\gevc.
Information from the five detector subsystems is combined in likelihood selectors which
 identify $e$, $\pi$, and $K$ particles and in a neural network which identifies muons.
The $\pi$-$K$ separation is provided by the DIRC and DCH whereas $\pi$-$\mu$ separation
is primarily accomplished with the IFR and EMC. 
 The identification efficiencies are given in Table~\ref{tablesyst} and
 cross-contaminations are given below.
We suppress di-muon and Bhabha backgrounds by requiring
signal tracks identified as a lepton to have \CM momentum less than 80\% of $\sqrt{s}/2c$. 
To reduce cross-feed from $e$ into the $\pi$ and $K$ channels,
the ratio of deposited electromagnetic energy of a $\pi$ or K candidate track to its measured momentum, $E/pc$, 
is required to be less than 0.85.
A pion track also passing a loose muon selection is rejected.
A similar veto is applied for a kaon track passing the loose muon selection if its measured momentum exceeds 3\gevc.
Also, events with an EMC energy $ > \{1.0, 0.5, 0.2,0.2 \}\gev$ 
in the signal hemisphere unassociated with the $\{e, \mu, \pi, K\}$ track are removed.

Pion and kaon control samples from $D^{*+}\ra \pi^+D^0, D^0\ra\pi^+K^-$ decays
are used to study and correct for small differences between MC and data. 
We cross-check these with independent
$\pi^+$ ($K^-$) control samples from $\tau^-\to \pi^-\pi^-\pi^+ \nu_{\tau}$ ($\tau^-\to K^-\pi^-K^+ \nu_{\tau}$)
decays using particle identification of two of the oppositely charged particles and 
the fact that the wrong sign $\tau^-\to \pi^- \pi^- K^+\nu_{\tau}$ decays are heavily suppressed.
Samples of radiative Bhabha and radiative $\mu$-pair events provide control samples of electrons and muons.
The systematic uncertainty associated with charged particle
identification is assessed from the control sample statistical
errors, consistency between control samples, and the sensitivity of
the control sample corrections to the number of particles near the track.
The statistical errors in the
more limited cross-check control samples dominate these errors. 
Because we use control samples to correct charge conjugate particles separately,
charge-dependent detector responses are accounted for by construction. 

To remove two-photon and Bhabha backgrounds, the event must have a missing \CM energy between 10\% and 70\% of $\sqrt{s}$. 
The angle between the missing momentum and electron beam direction in the \CM, \THETAMISSCM ,
is constrained to satisfy $|\cos(\THETAMISSCM)|<0.7$,
the thrust 
  of the event is required to be above 0.9,
and the net missing transverse momentum in the \CM  greater than 0.009$\sqrt{s}/c$. 

\begin{figure}[t]
\resizebox{0.5\textwidth}{0.45\textheight}{%
\includegraphics{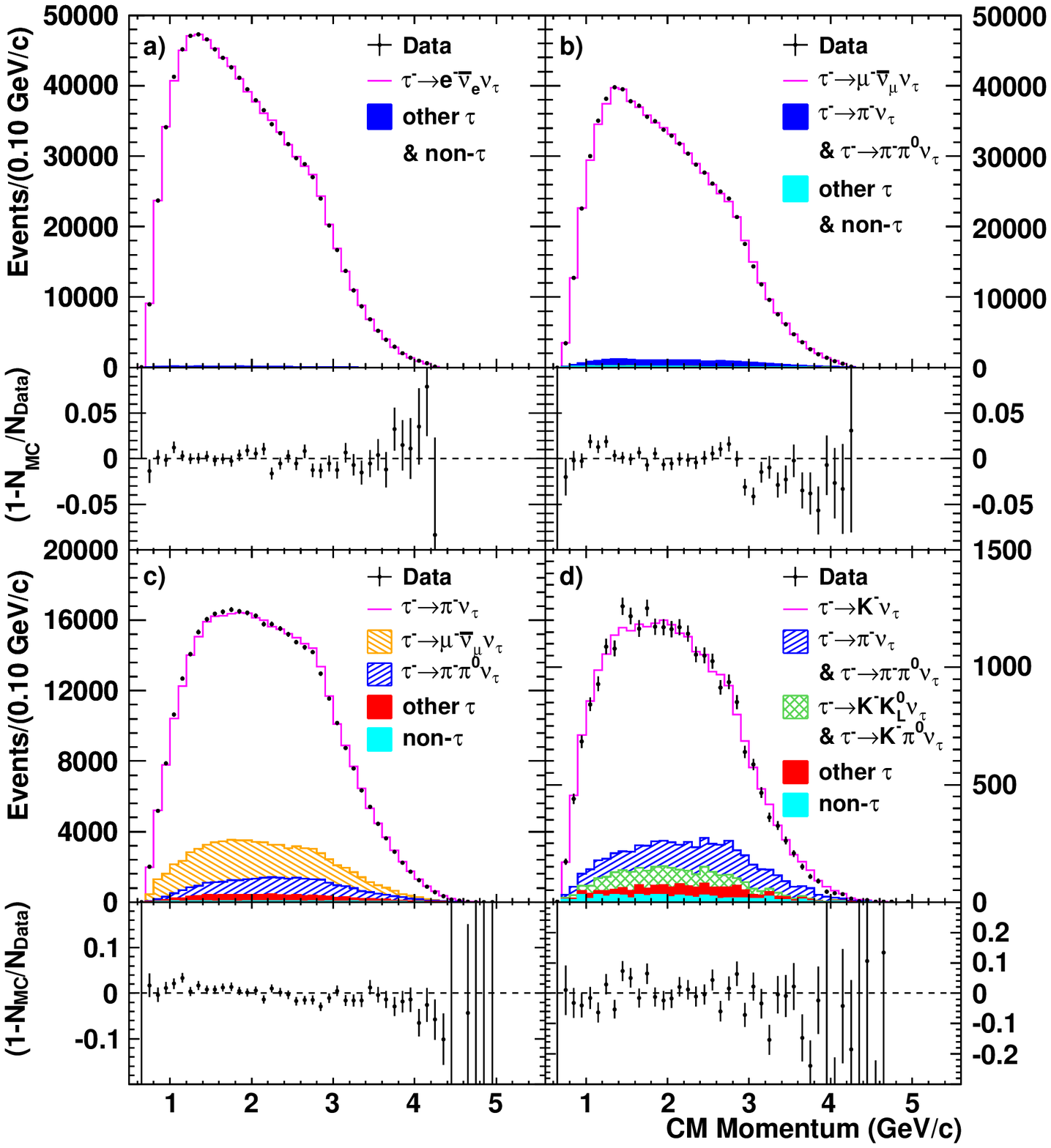}
}
\caption{Data (points) and MC (histograms) distributions of \CM momentum for (a) \tauenu, (b) \taumunu, (c) \taupinu
 and (d) \tauknu modes.
The small differences between MC and data are accounted for in the systematic errors.
}
\label{figurePcm}
\end{figure}

Each of the three tag-side tracks has an electron veto applied to further reduce
the Bhabha contamination. This results in less than 0.03\% contamination
from two-photon events and less than 0.1\% contamination from Bhabha
events in the electron signal sample.
These backgrounds were investigated by studying 
samples enriched  in  Bhabha and two-photon events by adjusting the
requirements on  the thrust, $\cos(\THETAMISSCM)$, and transverse
momentum of the event.
Potential background from Bhabha events were further probed by studying the number of events 
having a high signal track momentum as the electron veto was progressively lifted from one, 
then two, and finally all three tracks in the tag hemisphere.

To suppress backgrounds in  the $\taupinu$ and $\tauknu$ channels from $\tau$ decays
with undetected neutral particles  other than the  $\nu_{\tau}$ (e.g. $K^0_L$ mesons, $\nu_{\mu}$),
we reconstruct the direction of the back-to-back $\tau^+\tau^-$ system in the \CM frame.
The polar angle of the $\tau$ momentum with respect to the tag-side hadronic system 
is calculated assuming that the \CM energy of the $\tau$ is $\sqrt{s}/2$, and
the azimuthal angle of the $\tau$ momentum is fixed to a value that has been optimized to minimize
the total error on $\BR_{K/\pi}$~\cite{Nugent}. With this estimator for the $\tau$ momentum, we require
the missing mass in the signal hemisphere to be less than 0.56~\gevcc.

For the selected $\taumunu$ events, the dominant backgrounds  are
$\tau^{-}\to  \pi^{-}\nut$ $(1.46 \pm 0.01)\%$ and
$\tau^{-}\to  \pi^{-}\pi^{0}\nut$ $(0.85 \pm 0.01)\%$.
For the  $\taupinu$ channel, the dominant backgrounds  are
$\tau^{-}\to  \mu^{-}\numb \nut$ $(12.90 \pm 0.07)\%$,
$\tau^{-}\to  \pi^{-}\pi^{0}\nut$ $(5.87\pm 0.04)\%$,
and non-$\tau$ backgrounds $(0.34 \pm 0.05)\%$.
The major backgrounds in the $\tauknu$ channel are from
$\tau^{-}\to  \pi^{-}\nut$ decays $(10.06 \pm 0.13)\%$,
$\tau^{-}\to  K^{-}\KL\nut$ $(3.87 \pm 0.41)\%$,
$\tau^{-}\to  K^{-}\pi^{0}\nut$ $(1.97 \pm 0.14)\%$,
$\tau^{-}\to  \pi^{-}\pi^{0}\nut$ $(1.07 \pm 0.06)\%$,
and non-$\tau$ backgrounds $(2.58\pm 0.38)\%$.
The uncertainties are from MC statistics, branching fractions and, 
for non-$\tau$ backgrounds, the systematic uncertainty on background rates.
Fig.~\ref{figurePcm} shows the momentum distributions in the \CM frame for
each of the four decay modes for data,
along with the background MC contributions.

For the $\tauenu$ channel, \NDataenu events are selected with an efficiency and purity of \Effenu and \Purenu, respectively.
The number of selected events, efficiency, purity and systematic uncertainties on $R_i$ 
of the  $\taumunu$, $\taupinu$, and $\tauknu$ selections are presented in Table~\ref{tablesyst}.
These uncertainties include contributions from the particle identification, 
the sensitivity to detector response including 
the impact of changing the MC momentum scale and DCH resolution, 
modelling of hadronic and electromagnetic showers in the EMC, the EMC energy scale, 
and angular measurements made by these detectors within their modelling uncertainties,
the backgrounds, initial- and final-state radiation, radiation in $\tau$ decays, 
rate and shape of $\tau^{-}\to\pi^{-}\pi^{-}\pi^{+}\nut$ decays,
the trigger, and $\mathcal{L}\sigma_{\eett}$.
The systematic uncertainty on $R_\mu$ is dominated by uncertainties in  particle identification. 
The $R_\pi$ and $R_K$ measurements have additional dominant contributions from the detector modelling and associated backgrounds, 
due to stronger cuts on the EMC energy necessary to reduce non-$\tau$ backgrounds.
Presence of the $\sim$20\% backgrounds in these channels render them more sensitive to the modelling of the tag-side decays.
 The dominant background uncertainty in the $R_\pi$ measurement arises from the
electron contamination in the $\pi$ sample investigated by measuring
the number of events that fail the $E/p$ electron veto requirement
in data and MC. In the $R_K$ event sample, the uncertainty arising 
from $\tau$ decay branching fractions of background modes is 
0.58\%, which is dominated by the uncertainty of 
the $\tau^-\to K^0_L K^- \nu_{\tau}$ fraction.
There is also a 0.49\%  uncertainty assigned for $q\bar{q}$ backgrounds, which are
studied using events with an invariant mass of the tracks in the tag hemisphere above the $\tau$-mass
and cross-checked in regions of thrust and $\cos(\theta_{\rm{miss}}^{\rm{CM}})$
enriched with these backgrounds.

\begin{table}[h]
\caption{Number of selected events, purity, total efficiency, component of the efficiency from particle
identification,
 and systematic uncertainties (in \%) on $R_i$ for each decay mode.}
\begin{center}
\begin{tabular}{lccc} \hline
                                 &  $\mu$       & $\pi$      & $K$        \\ \hline
 $\mathbf{N}^{\rm{D}}$            &  \NDatamunu & \NDatapinu & \NDataknu  \\
 Purity                          &  \Purmunu   & \Purpinu   & \Purknu    \\
 Total Efficiency                &  \Effmunu   & \Effpinu   & \Effknu    \\ 
 Particle ID Efficiency          &  74.5\%       &  74.6\%      & 84.6\%    \\ \hline
\multicolumn{4}{l}{Systematic uncertainties:}\\ 
Particle ID                      & 0.32        &  0.51       &   0.94    \\
Detector response                & 0.08        &  0.64       &   0.54    \\
Backgrounds                      & 0.08        &  0.44       &   0.85    \\ 
Trigger                          & 0.10        &  0.10       &   0.10    \\
$\pi^{-}\pi^{-}\pi^{+}$ modelling & 0.01        &  0.07       &   0.27    \\ 
Radiation                        & 0.04        &  0.10       &   0.04    \\ 
$\BR(\tau^-\to\pi^{-}\pi^{-}\pi^{+}\nu_{\tau})$  
                                 & 0.05        &  0.15       &    0.40   \\ 
$\mathcal{L}\sigma_{\eett}$       & 0.02        &  0.39       &    0.20   \\ \hline
Total [\%]                       & 0.36        &  1.0        &    1.5    \\ \hline 
\end{tabular}
\label{tablesyst}
\end{center}
\end{table}
The measured branching ratios and fractions are:
\begin{eqnarray*}
R_\mu &=& \BRmunu\\
R_\pi &=& \BRpinu\\
R_K  &=& \BRknu \\
R_h = R_\pi + R_K  &=& \BRhnu \\
\BR(\taumunu) &=& \BFmunu\\
\BR(\taupinu) &=& \BFpinu\\
\BR(\tauknu)  &=& \BFknu
\end{eqnarray*}
where   $h$ = $\pi$ or $K$ and we use  $\BR(\tauenu)=(17.82\pm0.05)\%$~\cite{PDG}.
The off-diagonal elements of the correlation matrix for the measured ratios (branching fractions) are
$\rho_{\mu \pi}=$0.25 (0.34), $\rho_{\mu K}=$0.12 (0.20), and $\rho_{\pi K}=$0.33 (0.36).
The $\mu$ and $\pi$ measurements are consistent with and of comparable precision as
 the world averages~\cite{PDG} whereas the $K$ measurement is consistent with but twice as
 precise as the world average~\cite{PDG}.

Tests of $\mu-e$ universality can be expressed as
\begin{eqnarray*}
\left( \frac{g_\mu}{g_e} \right)^2_{\tau} = \BRtautomunu \frac{f(m_e^2/m_\tau^2)}{f(m_\mu^2/m_\tau^2)},
\end{eqnarray*}
where $f(x) = 1-8x+8x^3-x^4-12x^2\log{x}$, 
assuming that the neutrino masses are negligible~\cite{Tsai:1971vv}.
This gives $\left( \frac{g_\mu}{g_e} \right)_{\tau}$ = \gmgevalue,
yielding a new world average of $1.0018\pm0.0014$,
which is consistent with the SM and the value of $1.0021\pm0.0015$ from pion decays~\cite{PDG,Cirigliano:2007ga}.

Tau-muon universality is tested with
\begin{eqnarray*}
\left( \frac{g_{\tau}}{g_{\mu}} \right)^2_{h} = \frac{\BFtautohnunc}{\BFhmutwonc} \frac{2m_h m^2_{\mu}\tau_h}{(1+\delta_{h})m^3_{\tau}\tau_{\tau}} 
                                       \left( \frac{1-m^2_{\mu}/m^2_h}{1-m^2_h/m^2_{\tau}} \right)^2,
\end{eqnarray*}
\noindent where the radiative corrections are
$\delta_{\pi} = (0.16 \pm 0.14)\%$ and
$\delta_{K} = (0.90 \pm 0.22)\%$~\cite{Marciano:1993sh}.
Using the world averaged mass and lifetime values and meson decay rates~\cite{PDG},
we determine $\left( \frac{g_{\tau}}{g_{\mu}} \right)_{\pi(K)}$ = \gtgmpvalue (\gtgmkvalue) and
$\left( \frac{g_{\tau}}{g_{\mu}} \right)_{h}$ = \gtgmhvalue when combining these results;
this is 2.8$\sigma$ below the SM expectation and within 2$\sigma$ of the world average.

We use the  kaon decay constant $f_K = 157 \pm 2 \mev$~\cite{Follana:2007uv}, 
and our value of 
\begin{eqnarray*}
\BR(\tauknu) = \frac{G^2_F f^2_K \Vus^2 m^3_{\tau} \tau_{\tau}}{16\pi\hbar} \left (1 - \frac{m_K^2}{m_\tau^2} \right )^2 S_{EW},
\end{eqnarray*}
where $S_{EW} = 1.0201\pm0.0003$~\cite{Erler:2002mv}, to determine $\Vus= 0.2193 \pm 0.0032$.
This measurement is within 2$\sigma$ of the value 
of \Vusunit predicted by CKM unitarity  and is also
consistent with the value of  $\Vus = 0.2165 \pm 0.0027$ 
derived from the inclusive sum of strange $\tau$ decays~\cite{Maltman:2008ib}.

Both of our measured \Vus values depend on absolute strange decay rates. 
Our value of $R_{K/\pi}=\BRkpi$, however,  provides a \Vus value
driven by the ratio between strange and non-strange decays.
We  use $f_K/f_\pi = 1.189 \pm 0.007$~\cite{Follana:2007uv}, 
\Vud~\cite{Vud}, and
the long-distance correction $\delta_{LD} = (0.03\pm 0.44)\%$  estimated~\cite{Banerjee:2008hg} 
using corrections to $\tau\to h\nu$ and $h \to \mu\nu$~\cite{Marciano:1993sh,Marciano:2004uf}  in
\begin{eqnarray*}
R_{K/\pi} = \frac{f_K^2 |V_{us}|^2}{f_\pi^2 |V_{ud}|^2}
 \frac{\left( 1 -  \frac{m_K^2}{m_\tau^2} \right)^2}{\left( 1 -  \frac{m_\pi^2}{m_\tau^2} \right)^2} (1+\delta_{LD}),
\end{eqnarray*}
to obtain \Vus = \Vusvalue where short-distance electro-weak corrections cancel in this ratio.
This value is consistent with CKM unitarity~\cite{Vud}
and 2.5$\sigma$ higher than \Vus from the inclusive sum of strange $\tau$ decays.

We are grateful for the excellent luminosity and machine conditions
provided by our \pep2\ colleagues, 
and for the substantial dedicated effort from
the computing organizations that support \babar.
The collaborating institutions wish to thank 
SLAC for its support and kind hospitality. 
This work is supported by
DOE
and NSF (USA),
NSERC (Canada),
CEA and
CNRS-IN2P3
(France),
BMBF and DFG
(Germany),
INFN (Italy),
FOM (The Netherlands),
NFR (Norway),
MES (Russia),
MEC (Spain), and
STFC (United Kingdom). 
Individuals have received support from the
Marie Curie EIF (European Union) and
the A.~P.~Sloan Foundation.

\bibliographystyle{ieeetr}
\bibliography{references}

\begin{thebibliography}{99}


\bibitem{CKM}
  N.~Cabibbo,
  Phys.\ Rev.\ Lett.\  {\bf 10}, 531 (1963);
%
  M.~Kobayashi and T.~Maskawa,
  Prog.\ Theor.\ Phys.\  {\bf 49}, 652 (1973).


\bibitem{Pich:2008ni}
  A.~Pich,
  Nucl.\ Phys.\ Proc.\ Suppl.\  {\bf 181-182}, 300 (2008).


\bibitem{PDG}
  C.~Amsler {\it et al.}  (Particle Data Group),
  Phys.\ Lett.\  B {\bf 667}, 1 (2008);
%
  B.~Aubert {\it et al.}  (BABAR Collaboration),
  Phys.\ Rev.\ Lett.\  {\bf 100}, 011801 (2008).


\bibitem{:2008ct}
  F.~Ambrosino {\it et al.}  (KLOE Collaboration),
  JHEP {\bf 0804}, 059 (2008), $\Vus = 0.2237\pm 0.0013$;
M. Antonelli {\it et al.}
(FlaviaNet Kaon Group), arXiv:0801.1817 (2008).



\bibitem{Vud}
  J.~C.~Hardy and I.~S.~Towner,
  Phys.\ Rev.\  C {\bf 79}, 055502 (2009), \Vud = \Vudvalue.


\bibitem{Logan:2009uf}
  H.~E.~Logan and D.~MacLennan,
  Phys.\ Rev.\  D {\bf 79}, 115022 (2009);
  M.~Krawczyk and D.~Temes,
  Eur.\ Phys.\ J.\  C {\bf 44}, 435 (2005).


\bibitem{Dorsner:2009cu}
  I.~Dorsner, S.~Fajfer, J.~F.~Kamenik and N.~Kosnik,
  Phys.\ Lett.\  B {\bf 682}, 67 (2009).


\bibitem{Loinaz:2002ep}
  W.~Loinaz, N.~Okamura, T.~Takeuchi and L.~C.~R.~Wijewardhana,
  Phys.\ Rev.\  D {\bf 67}, 073012 (2003);
  A.~Czarnecki, W.~J.~Marciano and A.~Sirlin,
  Phys.\ Rev.\  D {\bf 70}, 093006 (2004);
  W.~J.~Marciano,
  PoS {\bf KAON}, 003 (2008).


\bibitem{Maltman:2008ib}
  E.~Gamiz, M.~Jamin, A.~Pich, J.~Prades and F.~Schwab,
  PoS {\bf KAON}, 008 (2008);
  K.~Maltman, C.~E.~Wolfe, S.~Banerjee, I.~Nugent, and J.~M.~Roney,
  Int.~J.~ Mod.~Phys. {\bf A23}, 3191 (2008).

\bibitem{conjugate}
Charge conjugate $\tau$ decays are implied throughout.

\bibitem{Follana:2007uv}
  E.~Follana, C.~T.~H.~Davies, G.~P.~Lepage and J.~Shigemitsu
  Phys.\ Rev.\ Lett.\  {\bf 100}, 062002 (2008).


\bibitem{Banerjee:2007is}
S.~Banerjee, B.~Pietrzyk, J.~M.~Roney, and Z.~Was,
  Phys.\ Rev.\  D {\bf 77}, 054012 (2008).

\bibitem{detector}
B.~Aubert {\it et al.}  (BABAR Collaboration),
\nima{479}, 1 (2002).

\bibitem{kk}
  S.~Jadach, B.~F.~L.~Ward and Z.~Was,
Comput.\ Phys.\ Commun.\  {\bf 130}, 260 (2000).



\bibitem{tauola}
S.~Jadach, Z.~Was, R.~Decker, and J.~H.~K\"uhn,
Comp.\ Phys.\ Comm.\  {\bf 76}, 361 (1993).

\bibitem{photos}
E.~Barberio and Z.~Was,
Comp.\ Phys.\ Comm.\  {\bf 79}, 291 (1994).

\bibitem{Lange:2001uf}
D.~J.~Lange,
\nima{462}, 152 (2001).

\bibitem{Sjostrand:1995iq}
T.~Sj\"ostrand,
Comp. Phys. Comm. {\bf 82}, 74 (1994).

\bibitem{geant}
 S.~Agostinelli {\it et al.} (GEANT4 Collaboration),
\nima{506}, 250 (2003).

\bibitem{Nugent}
I.~M.~Nugent,
University of Victoria Ph.D. Thesis (2008), SLAC-R-936.

%

\bibitem{Tsai:1971vv}
  Y.~S.~Tsai,
  Phys.\ Rev.\  D {\bf 4}, 2821 (1971)
  [Erratum-ibid.\  D {\bf 13}, 771 (1976)].


\bibitem{Cirigliano:2007ga}                                                    
  V.~Cirigliano and I.~Rosell,                                                 
  JHEP {\bf 0710}, 5 (2007).












\bibitem{Marciano:1993sh}
  W.~J.~Marciano and A.~Sirlin,
  Phys.\ Rev.\ Lett.\  {\bf 71}, 3629 (1993);
  R.~Decker and M.~Finkemeier,
  Nucl.\ Phys.\  B {\bf 438}, 17 (1995).
  R.~Decker and M.~Finkemeier,
  Phys.\ Lett.\  B {\bf 334} (1994) 199.



\bibitem{Erler:2002mv}
  J.~Erler,
  Rev.\ Mex.\ Fis.\  {\bf 50}, 200 (2004).

\bibitem{Banerjee:2008hg}
  S.~Banerjee,
   Proc. of ICHEP08, eConf C080730,  arXiv:0811.1429 (2008). 





\bibitem{Marciano:2004uf}
  W.~J.~Marciano,
  Phys.\ Rev.\ Lett.\  {\bf 93}, 231803 (2004).



\end{thebibliography}

\end{document}